\documentstyle[prb,aps,twocolumn,floats,epsfig,twoside]{revtex}
\input entete.tex
\def\be{\begin{equation}}
\def\ee{\end{equation}}
\def \bea {\begin{eqnarray}}
\def \eea {\end{eqnarray}}
\def\td{{\tilde\Delta_0}}
\def\dn{{\Delta_0}}
\def\dd{{\Delta}}
\def\e{{\rm e}}
\def\cC{{\cal C}}
\def\cE{{\cal E}}
\def\<{\langle}
\def\>{\rangle}
\def\vec #1{\mbox{\boldmath ${#1}$}}
\def\svec #1{\mbox{\boldmath ${\scriptstyle #1}$}}
\begin{document}
\twocolumn[\hsize\textwidth\columnwidth\hsize\csname@twocolumnfalse\endcsname

\title{Collective Phenomena in Defect Crystals}
\author{Reimer K\"uhn$^1$\cite{eku} and Alois W\"urger$^2$\cite{ewu}}

\address{$^1$Institut f\"ur Theoretische Physik, Universit\"at Heidelberg\\ 
Philosophenweg 19, 69120 Heidelberg, Germany\\
$^2$Centre de Physique Mol\'eculaire Optique et 
Hertzienne\footnote{Unit\'e Mixte de Recherche CNRS 5798},
Universit\'e Bordeaux 1\\351 cours de la Lib\'eration, F-33405 Talence cedex}

\maketitle
\begin{abstract}

We investigate effects of interactions between substitutional defects on 
the properties of defect crystals at low temperatures, where defect motion 
is governed by quantum effects. Both, thermal and dynamical properties are
considered. The influence of interactions on defect motion is described via 
a collective effect. Our treatment is semiclassical in the sense that we
analyze collective effects in a classical setting, and analyze the influence
on quantized defect motion only thereafter. Our theory describes a crossover 
to glassy behavior at sufficiently high defect concentration. Our approach
is meant to be general. For the sake of definiteness, we evaluate most of our
results with parameters appropriate for Li-doped KCl crystals.

\end{abstract}

%
]
\section{Introduction}

Interacting quantum impurities occur in various systems, such as spin 
glasses, magnetic systems with random couplings, and ionic crystals with 
substitutional defects. Spin glasses have been studied as a prominent example  
for systems far from thermal equilibrium. Though real spin glasses are of
quantum nature, most theoretical approaches rely on classical models. Work 
on true quantum spin glasses has been mainly directed towards elucidating
the nature of the phase transition to a frozen low-temperature state and at
studying the influence of quantum fluctuations on the order parameter 
(for a recent review, see Ref.~\CITE{Bha98}). Likewise with quantum 
ferromagnets.\cite{CDS96} Theoretical work has concentrated on the
quantum phase transition that occurs at $T=0$. In disordered 
quantum ferromagnets, additional interest stems from the fact that rare 
fluctuations of the disorder configurations leading to Griffiths singularities 
have a much stronger effect in the presence of quantum fluctuations than in 
the corresponding classical systems.\cite{Bha98}. Pertinent experiments
have been performed on the diluted dipolar Ising magnet 
LiHo$_c$Y$_{1-c}$F$_4$,\cite{ExpRM} which exhibits both, magnetic and spin 
glass order, depending on the concentration $c$ of Ho ions. Agreement between
experiment and theory for this system is, however, still not in a satisfactory 
state.

Early work on ionic crystals with substitutional impurities, such as lithium 
or cyanide in various alkali halides (KCl:Li, KBr:CN,...), focused on the 
thermal properties;\cite{Nar70,Bri75} experimental findings at high 
concentration indicated the relevance of the dipolar interactions\cite{Fio71}. 
More recently, sensitive echo measurements gave precise information on pairs of 
interacting impurities that occur at low doping\cite{Wei95}. At impurity 
concentrations of a few hundred ppm, the dipolar couplings destroy the 
coherent motion of the impurities\cite{Ens97}. Though a few questions would 
seem settled,\cite{Wue96} the role of interaction, especially the dynamical 
aspects, is still controversial.\cite{Bur94,Bur98,Wue94,Wue97} 

At still higher concentration, one obtains mixed crystals that show glassy 
behavior with respect to the rotational motion of the impurities, such as 
KBr$_{(1-c)}$(CN)$_c$\cite{Bir84} or (MF$_2$)$_{(1-c)}$(LF$_3$)$_c$, with 
M=Ca,Sr,Ba. For this latter example of mixed fluorite crystals,\cite{Tu99} 
the coupled density of states of the impurity degrees of freedom was deduced 
from Raman spectra, at concentrations $c=0.05-0.45$. 

In the present paper, we study the consequences of interactions between 
substitutional defects for the low $T$ properties of defect crystals. 
We are interested in describing the effects of interactions on quantized
defect motion, as well as in analyzing the crossover to glassy behavior
at sufficiently high defect concentration. Interactions  may be of an 
electric dipolar nature, or mediated by elastic strain fields.

We consider $N$ impurities that are randomly distributed on a lattice 
with $N_0$ sites. The defect concentration $c=N/N_0$ is usually much smaller 
than unity; the most interesting physics typically occurs in a range from 
$c=10^{-5}$ to a few per cent.  

The impurity Hamiltonian comprises a one-particle crystal field potential 
that is identical for each defect, and an interaction term that reflects 
the random configuration on the host lattice. The one-particle potential 
for a given impurity is written most conveniently in terms of local 
coordinates. For Li impurities in a KCl host, these coordinates could be 
chosen as ${\vec v}_i={\vec R}_i-{\vec R}_i^0$, where ${\vec R}_i$ is the 
impurity position at the lattice site ${\vec R}_i^0$. Tunneling arises from 
degenerate minima at off-center positions of the crystal field potential 
$G({\vec v}_i)$. The local displacement $|{\vec v}_i|$ is significantly 
smaller than the lattice constant. For other systems, the ${\vec v}_i$ might 
describe orientational degrees of freedom of a defect molecule.
   
In a dipole approximation, the interaction of impurities $i$ and $j$ is 
linear in the local variables ${\vec v}_i$ and ${\vec v}_j$ and involves a 
coupling parameter $J_{ij}$. It may arise from electric dipole moments 
${\vec p}_i = q {\vec v}_i$, or from  elastic quadrupole moments; in both 
cases the interaction parameter $J_{ij}$ is proportional to the inverse 
cube of the impurity distance  $r_{ij}=|{\vec R}_i^0-{\vec R}_j^0|$, that is
$J_{ij} \sim r_{ij}^{-3}$. 

In our investigation of collective effects we shall, in what follows, 
resort to a simplified description in terms of an effective {\em scalar\/} 
mean-field analysis that consists of the following approximations. (i) The 
dependence of $J_{ij}$ on the relative angles of ${\vec v}_i$, ${\vec v}_j$, 
and $({\vec R}_i^0-{\vec R}_j^0)$, is simplified to two possible signs. (ii) 
The local coordinates ${\vec v}_i$ are replaced by the one-dimensional scalar
variables $v_i$, the one-particle crystal field potential by $G(v_i)$, and 
the dipolar interaction by $J_{ij} v_i v_j$. In this scalar version, the 
impurity Hamiltonian takes the form
\begin{equation}\label{ham}
{\cal H} = \sum_i \frac{p_i^2}{2 m} + U_{\rm int}(\{v_i\})
\end{equation}
with
\begin{equation}\label{uint}
U_{\rm int}(\{v_i\}) = -\frac{1}{2} \sum_{i\ne j} J_{ij} v_i v_j + 
\sum_i G(v_i)\ .
\end{equation}
The distribution $P_0$ of dipolar couplings $J_{ij}$ can be evaluated for 
spacings $r_{ij}$ much larger than the lattice constant, i.e., for small 
impurity concentration $c\ll 1$. In a continuum approximation, the  
$J_{ij} \sim r_{ij}^{-3}$ scaling translates into
\be\label{P0j}
  P_0(J_{ij}) = {1\over2(N_0-1)} {J_{\rm max}\over J_{ij}^{2}},
\ee 
for $J_{\rm max}/N_0\le |J_{ij}|\le J_{\rm max}$. The factor ${1\over2}$ 
accounts for the two possible signs of $J_{ij}$. The upper bound $J_{\rm max}$ 
is determined by the interaction of closest neighbors,
\be\label{J2}
  J_{\rm max} =  \frac{3 q^2}{4\pi\epsilon\epsilon_0 r_{\rm min}^3} .
\ee
where we have used the form of the dipole moment $p_i = q v_i$. (Note 
that $r_{\rm min}$ is of the order of the lattice spacing.) Since we have 
replaced the dipolar interaction in three dimensions by the simpler expression 
$J_{ij}v_iv_j$, the numerical constant in (\ref{J2}) is to some extent 
arbitrary. Using a different scheme\cite{Wue97} we would obtain ${2/3}$ 
instead of the factor ${3/4\pi}$. Because of the $J_{ij} \sim r_{ij}^{-3}$
scaling, the lower bound $J_{\rm max}/N_0$ gives the coupling of 
impurities whose distance is of the order of the sample size. (iii) Lastly,
we neglect the distance dependence of the couplings and approximate the 
random-site character of the defect problem by a mean field model with 
randomly distributed all-to-all connections. In a further simplifying step, 
the $J_{ij}$-distribution is taken to be of the Gaussian form
\be
  P(J_{ij}) = \sqrt{N \over2\pi}{1\over J} 
                   \exp\left(- {N J_{ij}^2\over 2 J^2}\right),
\ee                 
with zero mean $\overline{J_{ij}} = 0$ and finite second moment, 
\be
  \overline{J_{ij}^2} = J^2/N\ ,
\ee
which is determined in such a way that it coincides with that of the actual 
distribution function $P_0(J_{ij})$, 
\be
  \int d J_{ij} J_{ij}^2 P_0(J_{ij}) = {1\over N_0} J_{\rm max}^2\ . 
\ee 
When putting $J^2/N=J_{\rm max}^2/N_0$ and using $c=N/N_0$, we obtain the 
scaling of the parameter $J$ with defect concentration $c$,
\be\label{J}
  J = \sqrt{c} J_{\rm max}\  .
\ee
The Gaussian distribution $P$ is roughly constant for $|J_{ij}|$ smaller 
than $J/\sqrt{N}$ and vanishes rapidly for higher values, whereas the more 
precise function $P_0$ shows a power law behavior. We repeat that, with 
(\ref{J}), their second moments are identical. 

Within our scalar approximation, we take crystal symmetries into account by
demanding that the one-particle potential satisfies the condition $G(v_i) =
G(-v_i)$. More specifically we choose $G(v_i)$ to exhibit degenerate minima 
at finite displacement $v_i=\pm a$. As the simplest choice of an on-site 
potential with these properties we use  
\begin{equation}
G(v) = g\,(v^2 - a^2)^2\ .
\label{gofv}
\end{equation}
The coupling constant $g$ is chosen in such a way that the tunnel splitting
$\Delta_0$ for impurities moving in an isolated double well of the form 
(\ref{gofv}) reproduces experimentally observed results for the lowest 
excitation energy of an isolated defect. E.g., in the case of $^7$Li in a
KCl host, one would require $\Delta_0 \simeq 1.1$ K, given a value of $a 
\simeq 0.7$\,\AA.

Due to the simplifications introduced above, the interaction energy 
(\ref{uint}) of the defect Hamiltonian has been made to resemble that of 
the SK spin-glass model,\cite{SK75} albeit one with continuous degrees 
of freedom rather than Ising spins. Models of this type have recently been 
proposed as candidates for describing low temperature anomalies of glassy
and amorphous systems. Indeed, it has been shown\cite{KuHo97,KuUr00} that 
a frustrated interaction of the form considered here is able to produce a 
potential energy landscape comprising an ensemble of single- and double well 
configurations -- the latter with a broad distribution of asymmetries 
and, in the translationally invariant case, also barrier heights, even if 
one starts out with single-site potentials $G(v)$ which are a-priori of 
single well form.

Our model of interacting impurities can be analyzed exactly within mean--field 
theory, i.e. a self-consistent representation of the interaction energy as a 
sum of effective single-site potential energies
\begin{equation}\label{mfdec}
U_{\rm int}(\{v_i\}) \longrightarrow \sum_i U_{\rm eff}(v_i)
\end{equation}
becomes exact in the thermodynamic limit. The ensemble of
effective single site potentials $U_{\rm eff}(v_i)$ represents the potential
energy landscape of the system of interacting defects. It will be found to 
contain {\em randomly\/} varying parameters whose distribution can be {\em 
computed}.\cite{KuHo97} By quantizing the defect motion within the collectively 
determined ensemble of effective single site potentials $U_{\rm eff}(v_i)$,
one finally obtains a semiclassical description of interaction effects on the 
behavior of quantum impurities.

We have organized the remainder of our material as follows. In Sec.~II we
analyze the potential energy landscape of the interacting system within 
mean-field theory. The analysis follows a proposal previously advocated for
glasses.\cite{KuHo97} Sec.~III describes  the analytic solution of the 
mean-field theory in the weak coupling regime. Thermodynamic consequences of 
interaction effects are explored in Sec.~IV, and simplifying features of 
two-state and WKB approximations are discussed in Sec.~V. Sec.~VI is devoted 
to dynamic effects, in particular to a computation of the distribution of 
relaxation rates, and to an analysis of the dynamic susceptibility. We 
discuss our results in some detail in Sec.~VII, comparing them with those 
of complementary approaches, and with experiments, and close  with a brief 
summary in Sec.~VIII.

\section{Mapping out the Potential Energy Surface}

To map out the potential energy surface of the interacting system,\cite{KuHo97}
one computes the configurational free energy
\bea
f_N(\beta) = -(\beta N)^{-1} \ln \int \prod_i d v_i 
\exp[-\beta U_{\rm pot}(\{v_i\})]\ ,
\eea
using replica theory to average over the ensemble of random $J_{ij}$ matrices,
so as to get {\em typical\/} results. The $T=0$ limit is eventually taken 
to select one of the (possibly many) classical ground-state configurations of 
the interacting system.
 
As announced above, a mean-field decoupling produces a collection of {\em 
independent single-site\/} potentials $U_{\rm eff}(v_i)$ with random 
parameters (which are comparable to randomly varying local fields in the 
context of spin-glasses). Technically, this decoupling within replica theory
can be seen as a method for the  self-consistent determination of the 
distribution of these random parameters.

We shall not repeat here the details of such a calculation, as they follow
standard lines of reasoning \cite{MPV87}. One obtains $f(\beta) = \lim_{n\to 
0} f_n(\beta)$ for the quenched free energy, with
\bea
n f_n(\beta)  &=&  \frac{1}{4}\beta J^2\sum_{a,b} q_{a,b}^2 \nonumber \\
& & - \beta^{-1} \ln \int \prod_a d v^a \exp\big[-\beta U_{\rm eff}(\{v^a\})
\big]\ .
\eea
Here
\begin{equation}\label{ueffn}
U_{\rm eff}(\{v^a\}) =  - \frac{1}{2}\beta J^2\sum_{a,b} 
q_{ab}v^a v^b + \sum_a G(v^a)
\end{equation}
is an effective replicated single-site potential, and the order parameters
$q_{ab} =N^{-1} \sum_i \langle v_i^a v_i^b\rangle$ are determined as solutions 
of the fixed point equations
\begin{equation}
q_{ab}  = \langle v^a v^b \rangle \quad ,\ a,b = 1,\dots ,n\ ,
\end{equation}
where angular brackets denote a Gibbs average corresponding to the effective
replica potential (\ref{ueffn}), and where it is understood that the limit 
$n\to 0$ is eventually to be taken.

We are, in what follows, going to solve the self-consistency equations only 
within the so-called replica symmetric (RS) ansatz for order parameters
\begin{equation}
q_{aa} = \hat q\ , \quad q_{ab} = q\ ,\quad a\ne b .
\end{equation}
The two order parameters of the RS ansatz must satisfy
\begin{equation}
\hat q  = \langle\, \langle v^2\rangle\, \rangle_z\ , \ \
q   =   \langle\, \langle v\rangle^2\, \rangle_z\ ,
\end{equation}
in which $\langle\dots\rangle_z$ denotes an average over a zero-mean 
unit-variance Gaussian $z$ while $\langle\dots\rangle$ without subscript is 
a Gibbs average generated by the effective replica-symmetric single--site 
potential
\begin{eqnarray}
U_{\rm RS}(v)  = - J \sqrt{q} z v -\frac{1}{2} J^2\cC v^2 + G(v)\ .
\label{urs}
\end{eqnarray}
with
$\cC=\beta (\hat q - q)$. The RS single site potential contains a Gaussian 
random variable $z$. A sum of single site potentials randomly drawn from
the RS Gaussian ensemble (\ref{urs}) constitutes the mean-field representation 
(\ref{mfdec}) of the potential energy landscape in terms of an ensemble of
effective single site potentials. Before analyzing this ensemble in greater
detail, let us note its two most salient features. 

First, due to a collective effect mediated by the interaction, there is a 
systematic deepening of the double-well crystal field potential experienced 
by the defects. Second, there is an induced distribution of asymmetries 
owing to the linear contribution to $U_{\rm RS}(v)$. Note that the deepening 
of the double well potential will in particular give rise to a renormalization 
of the tunnel splitting: $\dn \to \td < \dn$, reducing the smallest excitation 
energy that occurs in the system of tunneling impurities. However, due to 
the spectrum of asymmetries there will also be a spread of excitation energies 
towards larger values, as we shall describe in greater detail below.

Either of the fixed point equations for $\hat q$ or $q$ above may be replaced 
by one for $\cC=\beta (\hat q - q)$
\begin{eqnarray}
\cC = \frac{1}{J\sqrt q} \Big\langle\, \frac{d}{d z} \langle v\rangle\, 
\Big\rangle_z= \frac{1}{J\sqrt q} \Big\langle z \langle v \rangle
\Big\rangle_z\ \ ,
\end{eqnarray}
which turns out to aquire a finite value in the $\beta\to\infty$--limit that 
is of interest to us here. The RS free energy is
\bea
f_{\rm RS}& &(\beta) = \frac{1}{4} J^2 \cC (\hat q +q) \nonumber \\
& & - \beta^{-1} \left\langle\,\ln \int  d v 
\exp\big[-\beta U_{\rm RS}(v)\big]\right\rangle_z\ .
\eea

As the $\beta\to\infty$--limit is taken, Gibbs averages generated by $U_{\rm 
RS}(v)$ are dominated by the value(s) of $v$ which minimize $U_{\rm RS}(v)$,
which we denote by $\hat v = \hat v(z)$ (displaying its dependence on the
value of the Gaussian $z$). An immediate consequence is that $\hat q = q$ in 
this limit, provided that $q\ne 0$. The $T=0$ fixed point equations are then
\begin{equation}
\hat q = q =  \big\langle\, \hat v(z)^2\, \big\rangle_z\ \  ,\
\cC = \frac{1}{J\sqrt q} \Big\langle\, \frac{d}{d z} \hat v(z) 
\Big\rangle_z\ \end{equation}
with $\hat v= \hat v(z)$ minimizing $U_{\rm RS}(v)$, i.e., to be determined as
the appropriate solution(s) of
\begin{equation}\label{Gprime}
G'(\hat v) = J \sqrt{q} z  + J^2 \cC \hat v\ ,
\end{equation}
the prime denoting differentiation with respect to $v$. 
The solution $\hat v(z)$ of (\ref{Gprime}) is a smooth function of $z$, except 
at $z=0$ where $\hat v(z)$ has a jump-discontinuity, owing to the fact that 
among the solutions of (\ref{Gprime}) we have to choose the one corresponding 
to the {\em absolute\/} minimum of $U_{\rm RS}$. For $z \ne 0$ then, we obtain
\begin{equation}
\frac{d}{d z} \hat v(z) = \frac{J \sqrt{q}}{G''(\hat v) -  J^2 \cC}
\end{equation}
Taking the jump discontinuity of $\hat v(z)$ , hence the $\delta$--function 
singularity of its $z$--derivative at $z=0$ into account, we obtain the 
following form of the fixed point equation for $\cC$
\begin{equation}
\cC =  \Big\langle\, \frac{1}{G''(\hat v) -  J^2 \cC}\, \Big\rangle_z\ +
\frac{\Delta\hat v(0)}{J\sqrt {2 \pi q}}\ ,
\end{equation}
where $\Delta\hat v(0)$ denotes the size of the jump of $\hat v$ at $z=0$.
The $T=0$ limit of the free energy, i.e., the internal energy $u$ is
\begin{equation}
u = \frac{1}{4} J^2 \cC (\hat q +q) + \left\langle \, U_{\rm RS}(\hat 
v(z))\right\rangle_z\ .
\end{equation}
Owing to symmetry, the fixed point equations always admit of a $q=0$ solution. 
In a situation, where $G(v)$ is of double--well form (with minima at $\pm a$), 
however, only the $q\ne 0$ solution is thermodynamically acceptable as $T\to 
0$. Using the fixed point equations, we find that the internal energy is given
by
\begin{equation}
u = - J^2 \cC \hat q  + \left\langle \, G(\hat v(z))\right\rangle_z\ .
\end{equation}
in this limit. 
\section{The Low--Concentration or Weak--Coupling Regime}

The weak--coupling (low--concentration) limit is formally defined by the 
inequality $J a^2/g a^4 \ll 1$. It states that typical interaction energies 
are much smaller than the bare classical barrier height of the on--site
potential $G(v)$. In this limit, one has approximately $\hat v(z) \simeq 
\tilde a\, {\rm sgn}(z)$ for moderate values of the Gaussian $z$, hence 
$\hat q = q \simeq \tilde a^2$ and $\cC \simeq J^{-1}\sqrt{2/\pi}$. Here 
$\pm \tilde a$ denotes the coordinates of the minima of  $U_{\rm RS}(v)$ 
at $z=0$, that is 
\begin{equation}
\tilde a = a \left( 1+ \sqrt{\frac{2}{\pi}}\frac{J}{4 g a^2}\right)^{1/2}
\end{equation}
In the weak coupling limit, therefore, the effective single site potential 
experienced by the defects is of the form
\begin{equation}
U_{\rm RS}(v)  = - J \tilde a z\, v -\frac{J}{\sqrt{2\pi}} v^2 + G(v)\ ,
\label{urslc}
\end{equation}
exhibiting both, the systematic deepening of the double well potential 
experienced by the defects, and the induced distribution of asymmetries 
due to the linear contribution to $U_{\rm RS}(v)$ mentioned above. Both 
scale linearly with $J$, hence with the square root of the defect 
concentration in the low concentration regime (see (\ref{J})).

\begin{figure}
{\leavevmode
\centering
\epsfig{file=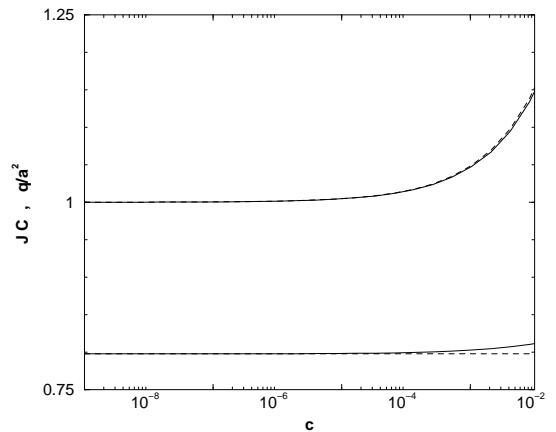,width=0.40\textwidth} 
\par}
\vspace{0.2truecm}
\caption[]{Test of low-concentration approximation. Upper set of curves: $q$,
lower set of curves: $J\cC$, given as functions of the defect concentration 
$c$. Dashed lines give the results of the low-concentration approximation, 
full curves, the numerical solutions of the full fixed point equations.}
\label{fig1}
\end{figure}

As Figure \ref{fig1}  shows, the results of the low concentration approximation 
agree fairly well with those of a full numerical solution of the fixed point 
equations for all concentrations of interest.

\section{Thermodynamics at Low Temperatures}

The contribution of the defects to the thermodynamics of the system at low
temperatures is dominated by quantum effects. That is, one has to consider the
quantized motion of the defects in the effective single site potentials given
by (\ref{urs}) or by their weak-coupling approximations (\ref{urslc}). 
The first task therefore consists in determining the energy levels $E_n$ and 
the corresponding eigenstates $\psi_n$ for the ensemble of effective single 
site Hamiltonians
\be
{\cal H}_{\rm eff} = \frac{p^2}{2 m} + U_{\rm RS} (v)\ .
\ee
This done, one may proceed to compute densities of state. Normalized with
respect to the total number $N_0$ of lattice sites, this gives
\be
\rho_n(E) = c\, \left\< \delta(E - \tilde E_n) \right\>_z \ ,
\ee
where $\tilde E_n = E_n - E_0$. The defect contribution to the specific 
heat in the same normalization gives
\be
C = c\, k_{\rm B} \beta^2 \left\< \large\<{\cal H}_{\rm eff}^2\large\>
- \large\<{\cal H}_{\rm eff}\large\>^2 \right\>_z\ ,
\ee
in which now
\be
\< \dots \> = \frac{{\rm Tr} (\dots ) \exp(-\beta {\cal H}_{\rm eff})}
{{\rm Tr} \exp(-\beta {\cal H}_{\rm eff})}\ .
\ee
Another quantity of interest is the static (dielectric) susceptibility,
originating from the interaction of the defect-dipoles with an external 
field $\cE$,  ${\cal H}_{\rm ext} = - q \cE\sum_i v_i$. The static 
Kubo-formula gives
\bea\label{susc}
\chi  =  c\, & & \beta q^2 \Bigg\< \Big\< v_{mm}^2 \Big\>
 - \Big\< v_{mm} \Big\>^2
\nonumber \\
  & & -\frac {1}{{\cal Z}_{\rm eff}} \sum_{m\ne n} \frac{e^{-\beta E_m}- 
      e^{-\beta E_n}}{\beta E_m - \beta E_n} v_{mn}^2\Bigg\>_z\ .
\eea 
The $v_{mn}=(\psi_n, v\,\psi_m)$ denote matrix elements of the position 
operator between the various eigenstates of the effective Hamiltonian 
${\cal H}_{\rm eff}$, and ${\cal Z}_{\rm eff}$ is the canonical partition 
sum generated by it.

\begin{figure}[h]
{\centering
\epsfig{file=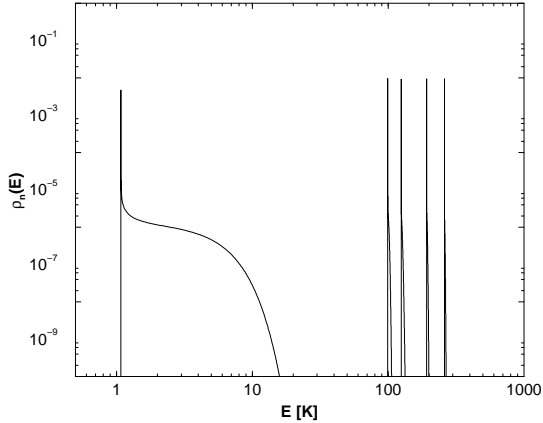,width=0.40\textwidth} 
\par}
\vspace{0.2truecm}
\caption[]{Density of states for the five lowest lying excitation energies at
$c=10$ ppm}
\label{fig2}
\end{figure}

As shown in Figure \ref{fig2}, the lowest band of excitation energies, 
originating from tunneling-excitations in the ensemble of asymmetric 
double-well potentials extends to much lower energies than the other
excitations. The latter are related to harmonic excitations about 
the two minima of $G$ with energies $\hbar\omega_0$ only beyond 
100 K. The pair of bands with peaks near 99\,K and 125\,K again 
corresponds to a pair of states mixed due to tunneling between the 
wells. The other two bands (with peaks near 192\,K and 260\,K correspond 
to states with energies above the barrier.

\begin{figure}[b]
{\centering
\epsfxsize=0.40\textwidth
\epsfig{file=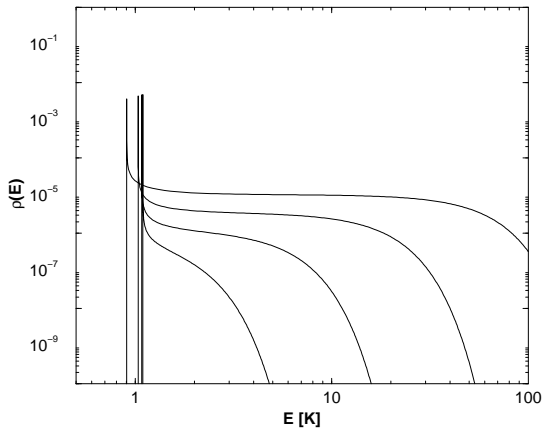,width=0.40\textwidth} 
\par}
\vspace{0.2truecm}
\caption[]{Density of states for the lowest band of tunneling excitations
at $c=$ 1, 10, 100, and 1000 ppm (bottom to top).}
\label{figdosc}
\end{figure}

\begin{figure}[h]
{\centering   
\epsfig{file=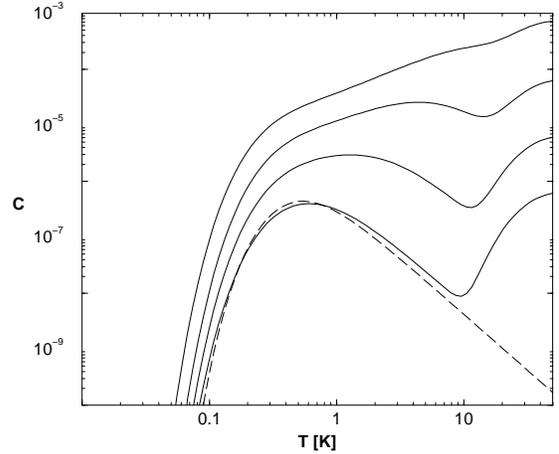,width=0.40\textwidth} 
\par}
\vspace{0.2truecm}
\caption[]{Specific heat as a function of temperature for $c=$ 1, 10, 100, 
and 1000 ppm (bottom to top). The influence of higher order excitations is 
seen only at temperatures above 10 K. The dashed curve is a Schottky-peak.}
\label{fig3}
\end{figure}

Tunneling excitations within the space spanned by the oscillator ground 
states in the two wells will dominate the low temperature physics. 
This is nicely seen in the specific heat data exhibited in 
Figure \ref{fig3}. Notice that the Schottky peak at low concentration 
is modified through interaction effects. As $c$ is increased beyond 
1000 ppm, a range of temperatures develops where the specific heat 
starts to show a linear temperature dependence much like in glasses. This
is due to the fact that the density of states corresponding to the (lowest
band of) tunneling excitations shown in Figure \ref{fig2} becomes nearly
constant in the energy range $\td<E<\sqrt{c}\, J_{\rm max}\, a^2$, which 
covers a large range of energies, as the concentration is sufficiently 
increased; see Figure \ref{figdosc} which also exhibits the renormalization 
of the tunneling matrix element towards lower energies due to interaction 
effects.

\begin{figure}[b] 
{\centering
\epsfig{file=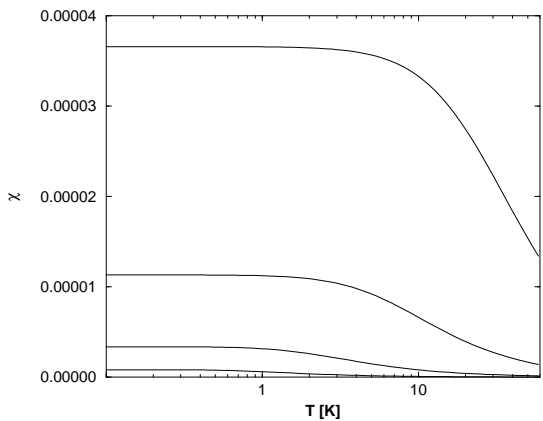,width=0.40\textwidth} 
\par}
\vspace{0.2truecm}
\caption[]{Static susceptibility  as a function of temperature for the same 
concentrations as in Figure 3.}
\label{fig4}
\end{figure}

In the static susceptibility plotted in Figure \ref{fig4}, no significant 
contribution of higher order excitations is detectable (on the scale of the
figure) at all, even up to temperatures as high as 60 K. The same quantity
evaluated in a two-state approximation would be indistinguishable from what 
is shown here. Note that the concentration dependence at low temperatures 
is proportional to  $c$ at low concentrations, but crosses over to a $\sqrt 
c$ behavior at larger concentrations. We shall return to this in greater 
detail later on.

\section{Two-state approximation}

At low temperatures, $k_{\rm B}T\ll\hbar\omega_0$, only the two lowest lying 
states are significantly populated.  For a sufficiently high barrier they can 
be constructed in terms of a basis of pocket states $|L\rangle$ and $|R\rangle$ 
that are localized about the potential minima at $\pm \tilde a$ of the two 
wells.

The corresponding Hamilton matrix reads
\be\label{e5-2}
  {\cal H} = {1\over2}\left(\matrix{ \dd  &  \td \cr
                                     \td &  - \dd }\right) ,
\ee                                     
where the off-diagonal elements are given by the renormalized tunnel energy.
The diagonal entries account for an asymmetry energy between the two wells,
$\dd=[U_{\rm RS}(-\tilde a)-U_{\rm RS}(\tilde  a)]$; with (\ref{urslc}) we find  
\be
\label{e5-4}
  \dd = 2 J \tilde a^2\, z  \equiv \bar\dd\, z\ .
\ee
The two-state Hamiltonian is easily diagonalized. Its eigenvalues are given by 
$\pm{1\over2}E$ with
\be\label{e5-6}
  E = \sqrt{\td^2+\dd^2};
\ee
the corresponding eigenstates $|\pm\rangle$ may be written in terms of a 
mixing angle $\tan\phi = \td/\dd$, 
\bea\label{e5-8}
  |+\rangle &=& \cos\phi\, |L\rangle + \sin\phi\, |R\rangle, \nonumber\\
  |-\rangle &=& \sin\phi\, |L\rangle - \cos\phi\, |R\rangle. 
\eea

Note that the asymmetry energy $\dd$ is linear in the Gaussian variable $z$; 
its distribution law is thus given by 
\be\label{e5-10}
  P_\dd(\dd) = {1\over\sqrt{2\pi}\bar\dd} \exp\left(-
{\dd^2\over2\bar\dd^2}\right) .
\ee
As a consequence, both the two-state energy splitting $E$ and the mixing angle 
$\phi$ are spread over a range determined by this distribution function. 

The resulting density of states for the lowest band of excitation energies
($\rho(E) = \rho_1(E)$ in the notation of the previous Section)
\be \label{e5-12}
  \rho(E) = c\, \left\< \delta\left(E-\sqrt{\td^2+\dd^2}\right)\right\>_z
\ee
is easily calculated. As is obvious from  (\ref{e5-6}), we have $\rho(E)=0$ 
for $E<\td$; at larger energies it is determined by the distribution of 
asymmetries, 
so
\be\label{e5-14}
  \rho(E) = c\,{2 E\over \sqrt{E^2-\td^2}}P_\dd \left(\sqrt{E^2-\td^2}\right)
\ee 
for $E\ge\td$, and shows a square root singularity for $E\to\td$; the 
pre-factor 2 is due the restriction $E>0$.

At very low concentration, the width of $P_\dd (\dd)$ is much smaller than the 
tunnel energy $\td$, and $\rho(E)$ is sharply peaked about $\td$. As soon as 
the typical interaction energy $\bar\dd \simeq 2 J\,a^2$ exceeds the tunnel 
energy $\td$, $\rho(E)$ is well approximated by a Gaussian above $\td$. It 
remains to compute the renormalized tunnel energy $\td$.

With respect to the computation of $\td$, we have looked at two variants. In
the first, both $\dn$ and its renormalized value $\td$ are computed exactly
by solving the Schr\"odinger equation of the impurity moving in the
appropriate crystal field potential, i.e.,  $G(v)$ for the former, and the 
renormalized potential $\tilde G(v) =  -\frac{J}{\sqrt{2\pi}} v^2 + G(v)$ for
the latter. As mentioned above, the first computation is actually used to fix
the coupling $g$ in (\ref{gofv}) so as to reproduce the experimentally observed
value -- e.g., $\dn \simeq 1.1$\,K for $^7$Li in KCl. With parameters 
appropriate for the KCl:$^7$Li example, this requires a potential with bare 
barrier height $g a^4 \simeq 174$\,K. 

In the second variant, both $\dn$ and its renormalized value $\td$ are 
computed within a WKB approximation. The bare tunneling matrix element is  
in this setting usually parameterized as
\begin{equation}\label{D0}
\Delta_0 = \hbar\omega_0 \exp(-\lambda) 
\end{equation}
with 
\begin{equation}\label{lamb}
\lambda = d\sqrt{2 m V_B/\hbar^2}\  ,
\end{equation}
and parameters derived from the bare on--site potential $G$ and the mass $m$
of the tunneling particle --- the frequency $\omega_0 = \sqrt{8 g a^2/m}$ 
of harmonic oscillations about the minima of $G$, the distance $d=2 a$ 
between the minima, and the height $V_B = g a^4 - \hbar\omega_0/2$ of the
classical barrier above ground state in the two wells. Once more, this first
calculation would be used to fix the coupling constant $g$ of the bare 
potential (\ref{gofv}). Within this approximation, the classical barrier 
required to reproduce the bare tunneling energy reported for KCl:$^7$Li 
is $g a^4 \simeq 84$\,K, i.e. only approximately {\em half\/} the value 
obtained from the numerically exact analysis. The {\em renormalized\/} 
tunneling matrix element is computed in the same way, except that parameters 
are computed from the renormalized on--site potential. This gives
\begin{eqnarray} \label{reno}
\tilde \omega_0 & = & \omega_0 \tilde a / a \nonumber\\
\tilde V_B & = & V_B + \frac{1}{2} \left(\sqrt{\frac{2}{\pi}} J a^2 + \hbar 
\omega_0 \Big(1 - \frac{\tilde a}{a}\Big)\right) \\
\tilde d & = & 2 \tilde a \nonumber
\end{eqnarray}
to lowest order in small parameters (with $\tilde a/a\simeq 1 + J/(4\sqrt{2 
\pi} g a^2$).

\begin{figure} 
{\centering
\epsfig{file=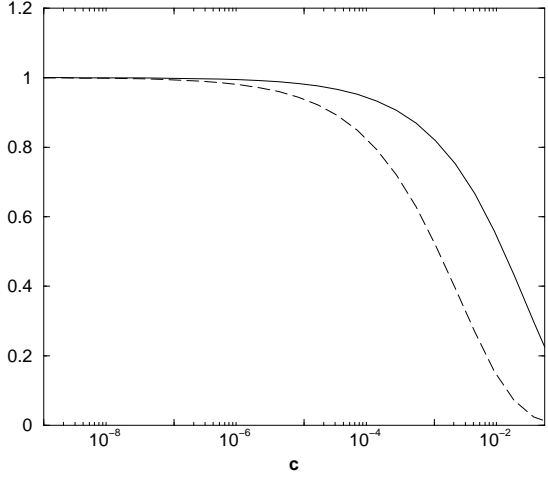,width=0.40\textwidth} 
\par}
\vspace{0.2truecm}
\caption[]{The ratio $\td/\Delta_0$ as a function of concentration. Exact
evaluation (full line), WKB approximation (dashed line).}
\label{figtdd}
\end{figure}

The renormalized tunnel energy $\td = \hbar\tilde\omega_0\e^{-\tilde\lambda}$, 
with 
\be
  \tilde\lambda=\tilde d\sqrt{2m\tilde V_B/\hbar^2}
\ee
may thus be expressed as
\be\label{td}
  \td = \dn \sqrt{1+\varepsilon}\,\e^{-\delta}\ .
\ee
Here
\be\label{eps}
\varepsilon=\sqrt{\frac{2}{\pi}}\frac{Ja^2}{4 ga^4}
\ee
and 
\be\label{delt}
\delta = \lambda\,\frac{\varepsilon}{2}\,\left(1 + 2\frac{ga^4}{V_B} - 
  \frac{\hbar\omega_0}{4 V_B}\right) + {\cal O}(\varepsilon^2)\ ,
\ee 
the latter quantity being evaluated in the weak coupling approximation, $J 
a^2\ll g a^4$ which is always appropriate for the case at hand. If, moreover, 
the barrier is assumed to be high, $\hbar\omega_0\ll V_B$, so that there are 
many oscillator levels between the potential minima at $v\simeq\pm a$ and the 
top of the barrier at $v=0$, the expression simplifies 
to 
\be
  \delta \simeq \lambda\frac{3\varepsilon}{2} = \lambda \frac{3}{8}
  \sqrt{\frac{2}{\pi}}\frac{Ja^2}{ga^4}\simeq \sqrt c\ 3 \sqrt{\frac{2}{\pi}}
  \frac{J_{\rm max} a^2}{\hbar \omega_0}\ ,
\ee 
in which corrections involving powers of $Ja^2/ga^4$ and $\hbar\omega_0/ga^4$
have been neglected. There is no restriction on the ratio $J a^2/\hbar\omega_0$.
It turns out, however, that in the KCl:Li case the ratio $\hbar\omega_0/ga^4$ 
is roughly 1.2, thus {\em not\/} small, so that the corresponding 
simplifications are not available. Figure \ref{figtdd} compares the ratio 
$\td/\dn$ evaluated numerically and via the WKB approximation (\ref{D0}),
(\ref{lamb}) and (\ref{td})-(\ref{delt}). The renormalization of the tunnel
energy becomes noticeable when the concentration exceeds 100 ppm, a 
concentration at which the typical asymmetry $\bar\dd$ becomes comparable 
with the vibrational energy $\hbar\omega_0$ in the two wells; the 
renormalization effect is, however, overestimated within the WKB approximation.

In the  two-state approximation, specific heat and static susceptibility are
given by
\be\label{sh2}
  C = c\,  k_{\rm B} \int {\rm d}E\rho(E) 
           {(\beta E/2)^2\over\cosh^2(\beta E/2)}. 
\ee  
and
\bea\label{susc2}
\chi  =  c\, \beta (q \tilde a)^2 \Bigg\< & & \frac{(\dd/E)^2} 
{\cosh^2(\beta E/2)}
\nonumber \\
  & & + 2\, k_{\rm B}T \frac{\td^2}{E^3}\, \tanh(\beta E/2)\Bigg\>_z\ ,
\eea 
respectively. The results are valid as long as temperature is much smaller 
than the librational energy $\hbar\omega_0$. In (\ref{susc2}) we have also 
introduced the usual approximate representation $v = \tilde a \sigma_z$ of 
the position operator in a basis of pocket states. This latter approximation 
is the largest source of errors in the expression (\ref{susc2}) for the 
susceptibility in that it overestimates matrix elements by roughly 10\%. 
It turns out tat this has a noticeable effect only on an overall pre-factor 
(at least for isolated or weakly coupled defects), but not on the 
temperature dependence. 

\section{Dynamics}

\subsection{Phonon damping}

The interaction of an impurity at site $i$ with elastic waves is described 
by the coupling potential
  \be
  \gamma \varepsilon v_i /a,
  \ee
with the elastic deformation potential $\gamma$, the reduced coordinate 
$v_i/a$, and the phonon strain field 
  \be
  \varepsilon = \sum_{{\svec q},s} \sqrt{\hbar \over 2 V\varrho 
  \omega_{{\svec q}s} }
                 q i\left(b_{{\svec q}s} - b^\dagger_{{\svec q}s}\right),
  \ee               
where $\varrho$ is the mass density of the host crystal, and ${\vec q}$ labels 
the wave vector of three acoustic phonon branches $s$.                
There is ample evidence that the defect-phonon coupling is weak; therefore 
it may be treated in first Born approximation. 

All dynamic information may be obtained from the two-time correlation 
function
  \be\label{ncorrel}
  G(t) = (1/2\<v^2\>)\<v(0)v(t)+v(t)v(0)\> ,
  \ee
with the normalization condition $G(0)=1$. At low temperatures, we may use 
the two-state approximation. We shall here restrict ourselves to the simplified 
representation $v=\tilde a\sigma_z$ of the position operator in the basis of
pocket states as introduced above; it is sufficiently precise to give 
qualitatively reliable results.

The theory of a weakly damped two-state system has been derived in many 
places; here we merely quote the result for the correlation function,
\bea\label{G}
  G(t) &=& {\td^2\over E^2} \cos(Et/\hbar)e^{-{1\over2}\Gamma t}\nonumber\\
       & & + {\dd^2\over E^2}\left((1-Q) e^{-\Gamma t} + Q\right),
\eea                      
with $Q=\tanh(\beta E/2)^2$ and the one-phonon damping rate
\be\label{Gam}
  \Gamma = {1\over2\pi} {3\gamma^2\over \hbar^4\varrho v_s^5}
               \td^2 E \coth(\beta E/2) .
\ee              
Here $v_s$ denotes the sound velocity. $G(t)$ determines the linear response 
of a two-state system with dipole moment $p=qa\sigma_z$, tunnel energy $\td$ 
and asymmetry $\dd$. There are several interesting issues arising from the 
above model. 

{\it First}, the Gaussian distribution for the bias $\dd$ leads to a 
distribution of resonance energies $E$ and relaxation rates $\Gamma$. 
  
{\it Second}, because of the variation of the typical asymmetry $\bar\dd$ 
with concentration $c$, the relaxation behavior of the impurities changes 
significantly with increasing $c$. From (\ref{G}) it is immediately clear 
that there is no significant relaxation contribution for $\bar\dd\ll\td$, 
i.e., at low concentration. In the opposite case $\bar\dd\gg \td$, the 
oscillatory (resonant) part of $G(t)$ is insignificant, and the relaxation 
term governs the impurity dynamics.  

{\it Third}, the temperature factor of the relaxation contribution, $1-Q=
\cosh(\beta E/2)^{-2}$, vanishes at very low temperatures $k_{\rm B}T\ll E$, 
whereas it tends towards unity at higher $T$ and large concentration.

\subsection{Rate distribution}

Recent measurements of the dielectric constant of KCL:Li revealed 
relaxational motion of the lithium impurities over several decades in the 
kHz range.\cite{Ens97} Such a broad relaxation spectrum is characteristic 
for disordered system in general. Therefore we discuss in some detail the 
distribution of relaxation rates 
  \be\label{e7-2}
  P_\Gamma(\Gamma) = \frac{1}{N} \sum_i \delta(\Gamma-\Gamma_i).
  \ee   
In our two-state model there is only one variable, $\dd=z\bar\Delta$, which 
is entirely determined by the Gaussian distribution of $z$ and the constant 
$\bar\Delta$; cf. (\ref{e5-6}). 

We give explicitly the rate distribution at zero temperature, because of its 
simplicity and since it shows the essential features. At $T=0$, Eq. (\ref{Gam})
gives
\be
\Gamma = \frac{\tilde\Gamma_0}{\td}\, E
\ee
with 
  \be\label{e7-6}
  \tilde \Gamma_0 = \frac{\td^3}{\dn^3}\Gamma_0 
   \hskip0.5cm {\rm and} \hskip0.5cm 
  \Gamma_0={1\over2\pi} {3\gamma^2\over \hbar^4\varrho v_s^5}\Delta_0^3\ .
\ee  
Hence, except for scales, the rate distribution is at the presently chosen
level of description basically equivalent to the density of states in the
two-level approximation. Substituting the Gaussian variable $z$ appearing in
the energy $E$, we find
\bea\label{e7-4}
P_\Gamma(\Gamma) &=& {1\over\sqrt{2\pi}}{\tilde\Delta_0\over\bar\Delta
	\tilde\Gamma_0}
             {2\,\Gamma\over \sqrt{\Gamma^2-\tilde\Gamma_0^2}}
             \nonumber\\
             & &\exp\left[-{1\over2}{\tilde\Delta_0^2 \over \bar\Delta^2
             \tilde\Gamma_0^2}
                 \left(\Gamma^2-\tilde\Gamma_0^2\right)\right] 
\eea
for $\Gamma\ge \tilde\Gamma_0$.

Here, $\tilde\Gamma_0$ is the minimum rate at finite doping. According to 
(\ref{e7-6}), it is reduced by a factor $\td^3/\dn^3$ as compared to the rate 
in the dilute limit, $\Gamma_0$. Like the density of states, $P_\Gamma(\Gamma)$ 
contains a Gaussian factor and a square root singularity at the minimum value 
$\tilde\Gamma_0$. The latter is of little relevance, and the rate distribution 
is governed the exponential in (\ref{e7-4}). We point out the most salient 
features of the rate distribution. 

\begin{figure} 
{\centering
\epsfig{file=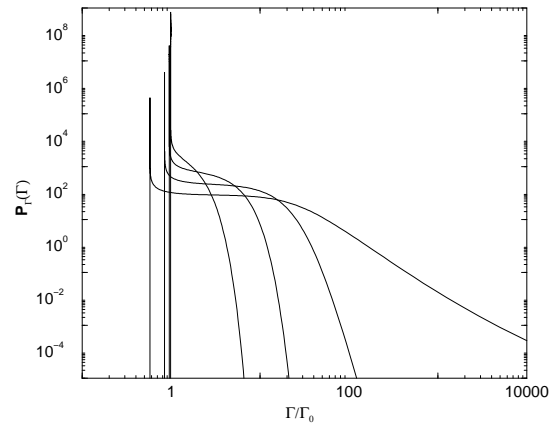,width=0.40\textwidth} 
\par}
\vspace{0.2truecm}
\caption[]{Rate distribution at $T=0$
for $c=$ 1, 10, 100 and 1000 ppm.}
\label{figrate}
\end{figure}

(i) For low impurity concentration, we have $\tilde\Delta_0 \gg \bar\Delta$, 
and $P_\Gamma(\Gamma)$ is a sharply peaked function with the lower cut-off 
$\tilde\Gamma_0$. In the dilute limit $c\to0$, the distribution tends towards 
a delta function at $\Gamma_0$. However, already at concentrations as low as
1 ppm, the rate distribution has a visible tail towards higher rates due to the
presence of asymmetries (see Figure \ref{figrate}).

(ii) In the opposite case of high doping, the maximum asymmetry exceeds by far 
the reduced tunnel energy, $\tilde\Delta_0 \ll \bar\Delta$. As a consequence, 
the Gaussian factor in (\ref{e7-4}) leads to a wide distribution that is almost 
constant between the lower bound $\tilde\Gamma_0$ and the effective upper 
cut-off $(\bar\Delta/\tilde\Delta_0)\tilde\Gamma_0$. 

(iii) The lower bound of the distribution, $\tilde\Gamma_0$, depends on the 
impurity concentration. In the dilute case, the tunnel energy is given by the 
bare value $\Delta_0$. Yet at finite concentration, the renormalization of the 
tunnel energy reduces the rate by the factor $\td^3/\dn^3$. At $c=0.01$, the 
minimum rate is by roughly one order of magnitude smaller than in the dilute 
case. (The WKB approximation predicts more than two orders of magnitude; 
see Figure \ref{figtdd}). Thus interaction leads to both slow and very fast 
relaxation, as compared to the low-doping case.

Figure \ref{figrate} shows the evolution from low to high doping. For 
concentrations up to 100 ppm, the curves have the same shape as those for 
the density of states (as predicted within the two-level approximation using
the pocket state approximation for the position operator. The curve for 1000 
ppm develops a different shape at high rates, because the pocket state
representation of the position operator becomes less and less precise at large 
asymmetries (compare Figures \ref{figrate} and \ref{figdosc}).

\subsection{Dynamic susceptibility}

Experimental investigations of the two-state dynamics in terms of elastic or 
dielectric response function involve the dynamical susceptibility $\chi(\omega) 
=\chi'(\omega)+ i\, \chi''(\omega)$, whose spectral function 
  \be\label{e7-10}
   \chi''(\omega) = c\, \frac{2}{\hbar} \tanh(\beta\hbar\omega/2)\,
   \langle G''(\omega)\rangle_z 
  \ee  
is related to the average of the motional spectrum $G''(\omega)$, i.e. of the 
Fourier transform of (\ref{G}). (The additional prefactor $(q\tilde a)^2$ 
appearing in the susceptibility (\ref{susc2}) and (\ref{susc}) is due to the 
fact that in Secs. III and IV we have considered an external field coupling 
to the {\em dipole operator\/} $q v$ rather than to a normalized position 
operator $v/\sqrt{\< v^2\>}$ whose correlator is considered in (\ref{ncorrel})).

Here we give explicitly the real part that describes the sound velocity or 
the reactive part of the dielectric function. For relevant frequencies 
$\hbar\omega\ll\td$ (and relaxation rates at typical concentrations 
satisfying $\Gamma_i\ll\td$) one has 
\bea 
\label{e7-12}
\chi'(\omega) &\simeq& {2\over N_0}\sum_i{\td^2\over E_i^3} \tanh(\beta E_i/2) 
\nonumber\\
& & + {\beta\over N_0} \sum_i {\dd_i^2\over E_i^2}\cosh(\beta E_i/2)^{-2} 
                 {\Gamma_i^2 \over \omega^2 + \Gamma_i^2} 
                 \eea
within the two-state approximation. There are two contributions of different 
origin to the susceptibility. The first or ``resonant" part is dominant in 
the dilute limit, where the typical asymmetry is significantly smaller than 
the tunnel energy, $\bar\Delta\ll\td$. On the other hand, the second 
contribution prevails at strong doping, where $\bar\Delta\gg\td$. 

\begin{figure} 
{\centering
\epsfig{file=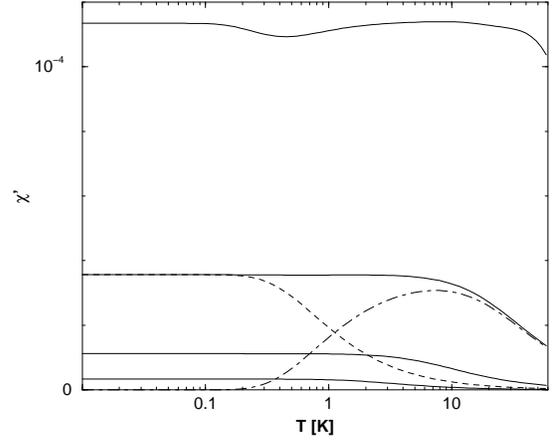,width=0.40\textwidth} 
\par}
\vspace{0.2truecm}
\caption[]{Real part of the susceptibility as a function of temperature at
$\omega= 10$\, kHz for $c$= 10, 100, 1000 and 10000 ppm (bottom to top). For
the $c=1000$\,ppm curve, the resonant (dashed) and relaxational (dot-dashed)
contribution are separately exhibited as well.}
\label{figrechi}
\end{figure}

In Fig. \ref{figrechi}  we plot $\chi'(\omega)$ as a function of temperature 
for various impurity concentrations $c$. At low doping, the susceptibility 
is of the van Vleck type and hardly depends on the external frequency. 
The temperature variation is given by the occupation difference of the 
two levels, $\tanh(\td/kT)$; accordingly $\chi'(\omega)$ is constant 
for $kT<\td$. At larger doping, pronounced relaxational contributions occur,
with peak-positions strongly dependent on frequency.

The zero-temperature value of $\chi'$ is a convenient measure of the 
relevance of interaction effects. The temperature factor in the 
relaxation contribution vanishes at $T=0$, and one has
\be \label{e7-14}
  \chi' = c\, \frac{1}{\td}\, \langle(1 + z^2 \bar\Delta^2/\td^2)^{-3/2}
  \rangle_z\ .
\ee
The average over $z$ can be written in terms of the confluent hyper-geometric 
function $U(a,b,x)$, 
  \be \label{e7-16}
  \chi'  = c {1\over\td} {1\over\sqrt{2\nu^2}}
                       U\left({1\over2},0,{1\over2\nu^2}\right), 
  \ee
where we have defined the dimensionless coupling parameter 
$\nu=(\bar\Delta/\td)$, 
  \be \label{e7-17}
   \nu = \sqrt{c}\,2\,{J_{\rm max}\tilde a^2\over \td} \simeq
   \sqrt{c}\,2\,{J_{\rm max}a^2\over \td} \equiv \sqrt{c\over c_0}.
  \ee
For weak doping, the asymmetry $\bar\Delta$ is much smaller than the 
tunnel energy; with $\nu\ll1$ and $U(1/2,0,1/2\nu^2)\approx\sqrt{2}\,\nu$ we 
recover the obvious result $\chi=c/\dn$. In the opposite case of strong 
doping we have $\nu\gg1$ and $U(1/2,0,1/\nu^2)=u_\infty=1.128...$ From 
(\ref{e7-16}) we thus obtain in the limiting cases 
  \be \label{e7-20}
  \chi'  = \cases{c /\td  & for $c\ll c_0$ \cr
              c/\bar\Delta= u_\infty \sqrt{c\,c_0}/\td & for $c\gg c_0$}. 
  \ee

The cross-over occurs at a concentration $c_0$ where the asymmetry attains 
the value of the tunnel energy, $\bar\Delta\approx\td$, or $\nu\approx1$. 
Eq. (\ref{e7-16}) yields
  \be\label{e7-18}
   c_0 = \left(\frac{\td}{2 J_{\rm max}a^2}\right)^2. 
  \ee  
Because of the concentration dependence of $\td$ this is an implicit equation 
for $c_0$. With parameters appropriate for $^7$KCl:Li as used before, one 
finds $c_0={\cal O}(10^{-6})$; compare Figure \ref{figchi_0_ofc}. 

The relaxation contribution to (\ref{e7-12}) strongly depends on the 
frequency $\omega$ and the rates $\Gamma_i$. Relaxation is most efficient 
where the rate of thermal two-level systems are close to the external 
frequency $\omega$; lowering the frequency shifts the peak to lower 
temperatures. The relaxation peak is significantly broadened by the 
rate distribution (\ref{e7-4}).

\section{Discussion}

\subsection{Cross-over to relaxation}

Broad spectra of energies and relaxation rates are characteristic for 
any disordered system. In the case of quantum impurities with dipolar 
interactions the broadening is tuned by the parameter $\nu$, which 
measures the strength of the interaction in units of the tunnel energy 
and increases with the square root of the impurity concentration $c$; 
cf. (\ref{e7-17}). 

At low doping the motional spectrum is peaked about the bare tunnel 
frequency $\dn/\hbar$. The relaxation rates that are close to $\Gamma_0$ 
as defined in (\ref{e7-6}) are of little importance, since the weight 
of the zero-frequency feature of the susceptibility is small. This changes 
as the impurity concentration reaches the value $c_0$, i.e. as $\nu$ 
tends towards unity. Then the average asymmetry energy is comparable to 
the tunnel energy; as a consequence, a strong relaxation peak emerges, 
and both the density of states $\rho(E)$ and the rate distribution 
$P_\Gamma(\Gamma)$ broaden; cf. Figs. \ref{figdosc} and \ref{figrate}.  

\begin{figure} 
{\centering
\epsfig{file=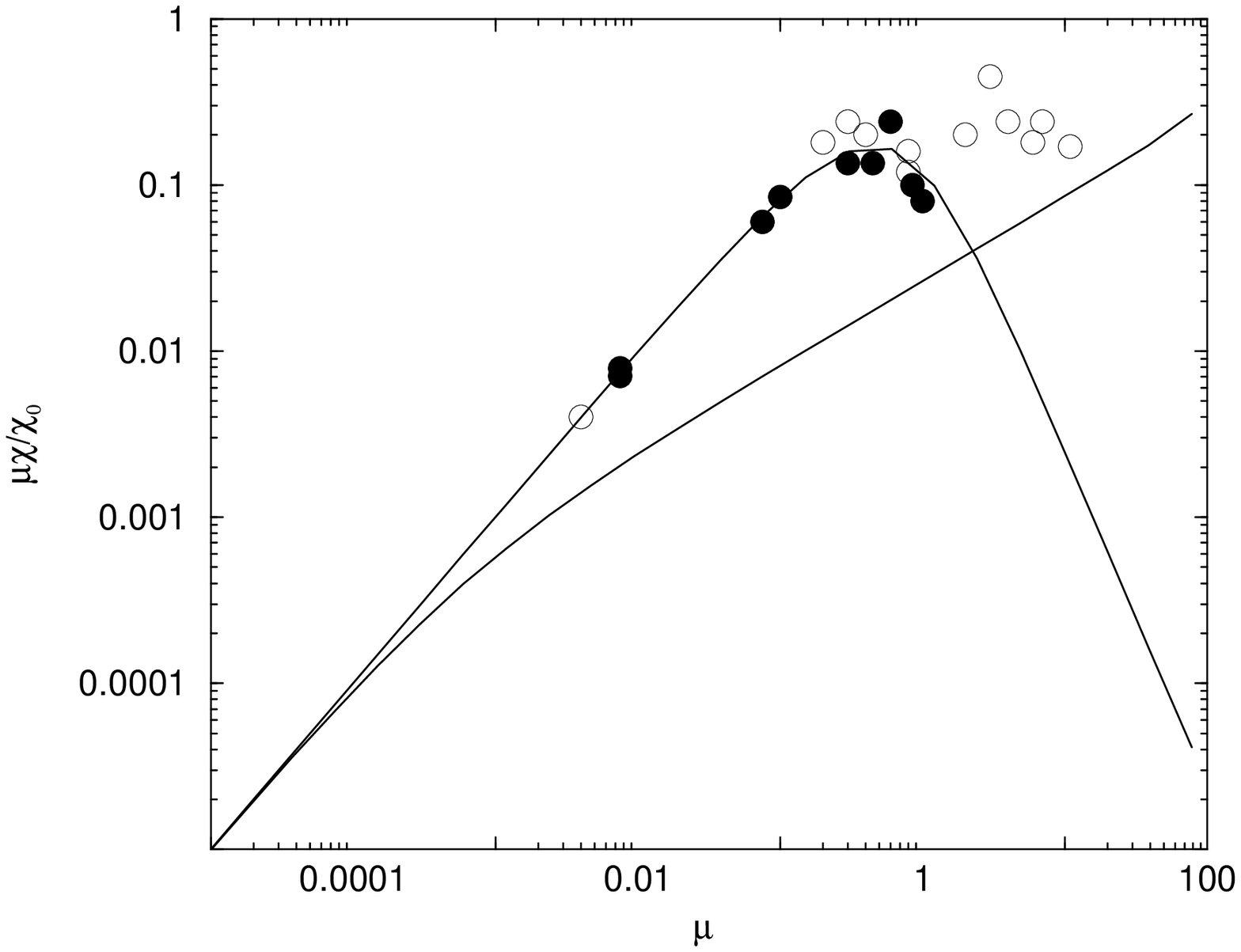,width=0.45\textwidth} 
\par}
\vspace{0.2truecm}
\caption[]{Zero temperature susceptibility as a function of the concentration. 
   The solid line gives the result of the present work, the dashed line that of 
   Ref. \CITE{Wue97}, i.e. Eq. (\ref{e8-2}), and the dotted line
   indicates non-interacting impurities. Full symbols indicate experimental 
   data on KCl:$^6$Li and KCl:$^7$Li [\CITE{Wei96,Wue96,Fio71}], open symbols 
   on KCl:OH [\CITE{Fio71,Moy83,Kae64}], LiF:OH [\CITE{Moy83}], RbCl:OH 
[\CITE{Fio71}], 
   NaCl:OH [\CITE{Moy83}]. (In order to obtain comparable quantities for these 
   various systems, we plot the rescaled susceptibility $\mu\chi/\chi_0$ as a 
   function of the dimensionless quantity $\mu$ that is linear in $c$.)}
\label{figchi_0_ofc}
\end{figure}

The static part of the susceptibility at $T=0$ provides the clearest
signature of this cross-over to relaxation. In Fig. \ref{figchi_0_ofc}
we compare the result of the present work (\ref{e7-16}) with that
obtained previously by one of us in a Mori projection scheme\cite{Wue97}, 
  \be \label{e8-2}
    \chi' = \chi_0 {(\sqrt{1+\mu^2} - \mu)^2 \over \sqrt{1+\mu^2}},
  \ee
where the dimensionless coupling constant 
  \be \label{e8-4}
    \mu = c\, 2\, {J_{\rm max} a^2\over\dn}
  \ee
may be considered as the rescaled concentration. (The extra factor 2 in
(\ref{e8-4}) as compared to the definition in Ref.~\CITE{Wue97} is due to
the fact that our convention for the coupling between impurities differs 
from that used in Ref.~\CITE{Wue97} by a factor 1/2). The susceptibility
in the absence of interaction, $\chi_0=c/\dn$, varies linearly with the
concentration; in Fig.  \ref{figchi_0_ofc} it is indicated as a
dotted line. Moreover we have indicated experimental data for various 
doped alkali halides. 
  
At very low doping, both expressions (\ref{e7-16}) and (\ref{e8-2}) are
linear in the concentration, as expected for non-interacting impurities,
whereas at higher impurity density the dipolar interactions of the tunnel
systems strongly reduce the susceptibility. We discuss the
differences of the theoretical results obtained from the present mean-field
model (\ref{e7-16}) and the Mori projection scheme (\ref{e8-2}). 

(i) Both parameters $\mu$ and $\nu$ are proportional to the ratio of the 
maximum interaction $J_{\rm max} a^2$ and the tunnel energy. In the present 
approach we found a square root dependence $\nu\propto\sqrt{c}$, whereas 
projection scheme yields a linear law $\mu\propto c$. (Note that in the 
expression for $\nu$ we have neglected the weak $c$-dependence of $\td$.)

(ii) As a consequence or these different powers, the cross-over to relaxation 
occurs at much higher concentration in (\ref{e8-2}), $\mu\approx1$ or 
$c_0=\dn/J_{\rm max} a^2$; the present mean-field theory yields, at 
$\nu\approx1$, the square of this quantity; cf. (\ref{e7-18}). 

(iii) In the strong-doping limit, the susceptibility in (\ref{e7-16}) varies 
as $\chi'\propto \sqrt{c}$; it still increases with concentration, whereas
(\ref{e8-2}) results in a decrease, $\chi'\propto c^{-2}$. 

These discrepancies are easily traced back to the basic assumptions of the 
models. The present mean-field approach relies on random couplings to 
{\em all\/} impurities. The characteristic interaction $J$ is given by the 
second moment of the Gaussian distribution and thus is proportional to the 
square root of the concentration; cf. (\ref{J}). 

In the previous Mori approach \cite{Wue97}, the characteristic 
interaction is determined by a few nearest neighbors of a given impurity. 
(Because of the $r^{-3}$-dependence on distance of the dipolar interaction 
and because of its random sign, neighbors at larger distances contribute 
little to field at a given site.) The $r^{-3}$-law results in the linear 
variation of $J$ with concentration. 

It should be noted that the square root law (\ref{J}) can be modified by 
a change of the mean-field model in a manner that takes the dominant effect 
of the interactions of a defect with a few nearest neighbors into account.
In a mean-field context this might be achieved  by replacing the fully 
connected random interaction matrix by a sparse random matrix such as in 
the Viana-Bray model\cite{VB85}, in which typical interaction scales may
be chosen to scale linearly with the concentration. In the details, however,
such a diluted mean-field model is much more complicated, requiring the 
introduction of infinitely many order parameters.

\subsection{Comparison with experiment}

The most significant result of the present work consists in the cross-over 
at $c=c_0$ from coherent one-particle motion to relaxation driven by the 
dipolar interaction. (The pre-factor $(\Delta/E)^2$ of the relaxation feature 
in the susceptibility (\ref{e7-12}) vanishes for zero interaction, whereas 
it tends towards unity in the strong-coupling limit.) 

Such a cross-over and the emergence of a relaxation peak in the motional 
spectrum have been observed for various impurity systems such as KCl:Li 
and NaCl:OH.\cite{Wue96} As discussed above and shown in Fig.
\ref{figchi_0_ofc}, the concentration dependence of the susceptibility in
the strong-doping limit obtained in the present work, $\chi'\propto\sqrt{c}$,
differs from the previous result $\chi'\propto c^{-2}$ (Ref. \CITE{Wue97}).
The data for the highest concentrations ($\mu>0.1$) clearly show the
relevance of the dipolar interactions; the measured values are orders of
magnitude smaller than expected for non-interacting impurities.

In spite of their marked differences of the theoretical laws (\ref{e7-16})
and (\ref{e8-2}), the available data do not permit us a definite statement
on the power law $\chi\propto c^\alpha$. Though (\ref{e8-2}) provides a
better fit at intermediate concentrations, $0.001<\mu<1$, the data
do not settle the power law at high concentrations. 
(Unfortunately, presently available crystals do not satisfy the strong 
doping criterion $\mu\gg1$.) Moreover, it is rather a difficult matter to
properly separate the relaxation contribution for strongly doped crystals;
thus the data points at the highest concentrations may well be too large.
Further experiments at high densities would be most desirable, in view of
the above discrepancy between $\alpha={1\over2}$ in the present approach and
$\alpha=-2$ from Ref. \CITE{Wue97}. 

We now turn to the rate distribution plotted in Fig. \ref{figrate}, that
shows a strong broadening with rising concentration. Indeed, the
relaxation spectra observed at high concentrations become very broad 
and develop a significant low-frequency wing \cite{Ens97}. On the other 
hand, specific heat measurements give clear evidence for a broadened density 
of states, in qualitative agreement with Fig. \ref{figdosc}. (Cf. Ref.
\CITE{Wue97} and original literature cited therein.) 

We close the discussion of alkali halides with a remark on the temperature
dependence of the relaxation amplitude in the susceptibility (\ref{e7-12}).
Since our mean-field theory reduces to a set of effective two-state systems,
the relaxation feature shows the well-known factor $\cosh({1\over2}\beta 
E)^{-2}$ and thus vanishes exponentially in the limit $T\to0$, whereas the 
rate $\Gamma$ tends towards a constant. This behavior is characteristic 
for a local degree of freedom. The Mori projection approach, on the other 
hand, results in a constant relaxation amplitude and a rate that decreases 
at low $T$, thus showing the generic behavior of collective relaxation. 
Again, the experimental situation would seem not entirely conclusive, though 
the data on KCl:Li of Ref. \CITE{Ens97} would suggest that both behaviors 
are present, i.e. the relaxation spectrum would comprise one-particle and 
collective contributions. 

Finally we discuss recent Raman light scattering experiments on mixed fluorite
crystals (MF$_2$)$_{(1-c)}$(LF$_3$)$_c$, with M=Ca,Sr,Ba.\cite{Tu99} The
reported data give strong evidence that the coupled density of states $D(E)$
of the two-level systems increases linearly with $c$ in the range $c=0.05-0.5$, 
i.e. $D(E)=c\, D_0(E)$; it would seem that the shape function $D_0(E)$ does not
change in this range. 

We consider very likely that the broad distribution of two-level energies 
observed in Ref. \CITE{Tu99} arises from elastic coupling of LF$_3$ tunneling 
states. In our model the TLS coupled density of states of is given by
  \be
  D(E)= {\tilde\Delta_0^2\over E^2} \rho(E).
  \ee
With the expression (\ref{e5-14}) for $\rho(E)$ one easily finds that 
$D(E)$ varies over a wide range as $E^{-2}$ and increases with $c$. Though 
these dependencies on $E$ and $c$ qualitatively agree with the data of Ref. 
\CITE{Tu99}, this experiment certainly requires a more careful study of the 
high-doping case. Possible discrepancies with the present description for 
this particular case arise from the manner in which defect crystals can,
or cannot become glassy at high doping, an issue to which we now turn.

\subsection{Crossover to glassy behavior}

The appearance in our theory of a broad spectrum of excitation energies 
and a correspondingly broad spectrum of relaxation rates due to interaction 
effects at high doping is reminiscent of the physics characteristic of
structural glasses. In this sense, our mean-field approach appears to
describe a crossover to glassy physics at high doping. Indeed, one of 
the original motivations for studying defect crystals has been that they 
constitute systems in which -- unlike in glasses -- the nature of the 
tunneling excitations is well understood, and which at the same time
offer the possibility of approaching a glassy limit by increasing the
defect concentration (for a recent review, see Ref. \CITE{Po+99}). It
should, however, be noted that the type of solely interaction-mediated 
glassiness described in the present paper is different from that 
believed to describe glasses proper in one essential aspect: while 
tunneling systems do occur with a broad distribution of asymmetries,
there is -- unlike in glasses -- {\em no\/} corresponding randomness 
in barrier heights. This feature entails\cite{HoKu99} for instance that 
the typical glassy plateau of the internal friction as a function of 
temperature would be absent in systems to which the present theory 
applies. On the other hand, such plateaus are known develop in certain
defect crystals such as (CaF$_2$)$_{(1-c)}$(LF$_3$)$_c$, but they require 
defect concentrations exceeding the 10\,\% range. There are basically two 
ways to create the required randomness also in the barrier heights. First, 
it {\em can\/} be solely interaction-mediated, if interactions are 
translationally invariant.\cite{KuUr00} This mechanism is, however, not 
available for a system of defects embedded in a crystalline host. The other 
possibility is to allow the crystal field potentials $G(v_i)$ to vary 
randomly from defect-site to defect-site. For highly doped defect crystals, 
this would seem like a realistic, and indeed expected feature. Unfortunately, 
though, there are no good theoretical models around to predict the associated 
kind of randomness and, in particular, its variation with defect 
concentration. In this sense, glassy defect crystals are apparently not much 
simpler systems than structural glasses proper.

\subsection{Comparison with classical reorientation model}

The dissipation rate (\ref{Gam}) describes jumps between two quantum levels; 
it may be decomposed as $\Gamma=\Gamma_\uparrow+\Gamma_\downarrow$, where
$\Gamma_\uparrow$ accounts for thermally activated jumps from the ground 
state upwards and $\Gamma_\downarrow$ for the reverse process. Detailed 
balance requires $\Gamma_\uparrow=e^{-\beta E}\Gamma_\downarrow$; thus we 
obtain with $E\gg kT$ the thermally activated rate
  \be\label{Gamclass}
  \Gamma_\uparrow = \tilde\Gamma_0 e^{-\beta E} ,
  \ee 
where the activation energy is, in     the strong-coupling limit, given by the 
dipolar interaction, $E=2 J \tilde a^2$. Such a rate, with an appropriate 
distribution of barrier heights, has been used in previous work on a 
classical reorientation model for quadrupolar glasses such as 
KBr$_{(1-c)}$(CN)$_c$ \cite{Bir84,Kan86}; in the classical picture the 
reorientation of the cyanide molecules (i.e. rotation by $\pi$) requires 
to overcome their quadrupole-quadrupole interaction that corresponds to 
our $2 J \tilde a^2$. Thus the present quantum mechanical calculation yields 
in the classical limit the proper temperature dependence of the rates.
(The present rate distribution is, however, much simpler than that derived 
in Ref.~\CITE{Kan86}.)

\subsection{Assumptions, Approximations and Potential Extensions}

Here we briefly summarize once more the  main assumptions and 
approximations made in this paper, trying to assess their quality,
and we indicate possibilities for further improvements.

(i) The 3D local coordinates of the impurities have been replaced 
by a scalar variable, and the dipolar couplings by scalar couplings 
with a random sign. In the analysis of dynamical effects we have
introduced a further simplification by truncating the Hilbert space of
the single-particle effective Hamiltonian to the ground state doublet. 
Whereas the neglect of two additional spatial coordinates should be 
of little relevance, in that the simplified model still brings out 
the salient features of the problem of interacting tunneling 
impurities, the truncation of the one-particle Hilbert space 
certainly breaks down as soon as the thermal energy or the mean dipolar 
interaction exceed the vibrational energy within one well of the 
impurity crystal potential. Yet this is not a severe constraint for 
our model since the most interesting physics occurs at low temperatures 
and for dipolar couplings comparable to the tunnel energy.

(ii) Interactions between tunneling impurities are analyzed classically
at the mean field level. This by itself is expected to give a reliable 
picture for much of the physics, as long as we are not interested in
coherent quantum motion of {\em several\/} particles (see below) or 
critical behaviour associated with phase transitions.

(iii) In actually performing the mean-field analysis we have replaced 
the distribution (3) by the Gaussian (5) with the same variance. As 
discussed in Sect. VII-A above, this is certainly a most serious 
approximation, which is dictated solely by the demand for analytic 
tractability, and which can not be justified on physical grounds. 
There would be basically two possibilities to improve on this within
random-bond modeling and replica techniques as used in the present paper.
One is to consider sparse connectivity models\cite{VB85}, the other
to follow a cavity approach proposed by Cizeau and Bouchaud\cite{CiBou93}. 
However, the former gives rise to an infinite set of order parameters and 
is quantitatively more or less untractable, the latter uses the two-state
nature of Ising spins in ways so essential to the analysis that we have 
seen no way to apply it to continuous degrees of freedom as they occur in
the present problem. Whether techniques invented for models with 
site-randomness will offer a way out of this dilemma, is currently under
study. 

Note that the uncertainties with respect to the concentration dependence 
alluded to above are quite general and arise in related approaches to, e.g., 
quantum spin sytems \cite{Gre97}.

A possible and uncontrolled ad-hoc modification of the theory which would 
give a $c^{1/3}$ behavior of the $T=0$ susceptibility at large doping 
instead of the $\sqrt c$ behavior, and which would therefore fit the high 
doping behaviour of this quantity rather well, is briefly mentioned in our 
summary below.

(iv) The approximation of replica symmetry used in the quantitative analysis
of collective behaviour is not expected to be a serious source of errors.
We have checked elsewhere\cite{KuHo97,KuUr00} for related systems that 
replica symmetry breaking effects are small e.g. in the low-temperature 
specific heat, but also in distributions of parameters characterizing 
double well-potentials.

(vi) The mean-field model itself constitutes an approximation that is by 
no means innocuous for {\em all\/} questions one might wish to look at. 
By mapping the pair interaction on an effective one-particle potential, 
we discard all effects of coherent motion of the impurities. 
As a straightforward consequence, the relaxation amplitude of the 
susceptibility (60) vanishes in the limit of zero temperature.

(vii) Finally, we briefly mention minor approximations that are of
little significance with respect to the basic features of our model.
In order to obtain simple laws in terms of the concentrations we
have truncated various powers series in Sect. VI, even if
higher-order corrections are not always negligible. 

\section{Summary}

We have studied interacting quantum impurities in terms of a mean-field 
model with a Gaussian distribution for the couplings. We briefly summarize 
our main results. 

(i) The dipolar interaction leads to a reduction of the tunneling 
amplitude and to a wide distribution of the asymmetry energy. As a 
consequence, the density of states is smeared out to both to smaller 
and higher values as compared to the unperturbed tunnel energy $\Delta_0$. 
Similarly, in the strong-coupling limit the relaxation spectrum covers 
several orders of magnitude. These features account for the collective
nature of the underlying relaxation process; they are in qualitative
agreement with experiments on lithium doped potassium chloride.\cite{Ens97} 

(ii) As a clear signature of interaction effects we consider how the 
zero-frequency susceptibility at low $T$ varies with the impurity 
concentration $c$. The present result, $\chi\propto\sqrt{c}$, strongly
deviates from the law $\chi\propto c^{-2}$ found previously in a different
approach. Though more recent experiments on KCl:Li would seem to speak in
favor of this latter result,\cite{Wue96} the data presently available in
the strong-doping regime $\mu>1$ do not unambiguously answer this question. 
Note, however, that the $\sqrt{c}$ scaling is clearly a consequence of the
`naive' mean-field  assumption of identically distributed all-to-all 
interactions. It is likely to be modified in more realistic approaches. 
For instance, by introducing an ad-hoc $c$ scaling of our interaction 
parameters in an otherwise structurally unmodified mean-field approach, 
one would obtain a $c^{1/3}$ behavior of the $T=0$ susceptibility at 
large doping instead of the $\sqrt c$ behavior.

(iii) Since the above square root behavior results from a generic feature 
of the Gaussian mean-field model, an experimental test of this law would 
be interesting for a wider class of models. Mean-field models with 
random couplings similar to that considered in this paper are frequently 
studied in view of quantum spin glasses and other disorder quantum 
systems.\cite{Gre97}

(iv) Our theory describes a crossover to glassy behavior in the (restricted)
sense that broad and rather flat distributions of excitation energies 
and relaxation times develop at high doping. We have argued that {\em true\/}
glassiness will develop in defect crystals only, when mechanisms are invoked
which are beyond those originating from defect interactions. As a consequence,
true glassy defect crystals are almost as difficult to grasp theoretically as
truly amorphous systems.

\acknowledgements

Very useful discussions with C. Enss, J. Classen, S. Hunklinger, S. Ludwig,
P. Nalbach, R.O. Pohl, M. Thesen, and B. Thimmel are gratefully acknowledged.
We dedicate this paper to Franz Wegner on the occasion of his 60-th birthday,
thanking him for numerous discussions and inspiration throughout the years.

\end{document}